\documentstyle[11pt,twoside,jltp,epsfig]{article}

\title{Relation between Vortex Pinning Energy and Anderson's Theorem}

\author{Nobuhiko Hayashi\address{Computer Center, Okayama University,
Okayama 700-8530, Japan} and Yusuke Kato\address{Department of Basic Science,
University of Tokyo, Tokyo 153-8902, Japan}}

\runninghead{N. Hayashi and Y. Kato}{Relation between Vortex Pinning
and Anderson's Theorem}

\begin{document}

\begin{abstract}
   We discuss the elementary vortex pinning in
type-II superconductors
in connection with the Anderson's theorem for nonmagnetic impurities.
   We address the following two issues.
   One is an enhancement of the vortex pinning energy
in the unconventional superconductors.
   This enhancement comes from
the pair-breaking effect of a nonmagnetic defect
as the pinning center far away from the vortex core
(i.e., the pair-breaking effect due to
the non-applicability of the Anderson's theorem
in the unconventional superconductors).
   The other is an effect of the chirality on the vortex pinning energy
in a chiral $p$-wave superconductor.
   The vortex pinning energy depends on the chirality.
   This is related to the cancellation of
the angular momentum between the vorticity and chirality in a chiral
$p$-wave vortex core,
resulting in local applicability of the Anderson's theorem
(or local recovery of the Anderson's theorem) inside the vortex core.


\end{abstract}

\maketitle


\section{INTRODUCTION}
   Much attention has been focused on the vortex pinning
in type-II superconductors.
   The vortex pinning governs the macroscopic magnetic properties of
type-II superconductors such as the hysteresis of the magnetization
and the critical current. The elementary vortex pinning potential is
the interaction between a vortex and a single defect.
   A microscopic analysis of the elementary vortex pinning potential is
necessary for understanding of the macroscopic vortex pinning problem in the
superconductors.~\cite{Thuneberg82,Thuneberg84,Thuneberg84:JLTP,Kerchner83,Kes}

   The vortex pinning energy is given by the difference in
the free energy
between the case when the vortex is located at the pinning center
and
the case when the vortex is far away from the pinning center.
   Naively, the elementary vortex pinning energy due to
a defect
with the volume $V_{\rm i}$ is expected to be given roughly by
\begin{equation}
\frac{1}{8\pi} p V_{\rm i} H^2_{\rm c},
\label{eq:conventional}
\end{equation}
where $H_{\rm c}$ is the thermodynamic critical field and
$p$ is a numerical factor much smaller than unity.
   The superconductivity is destroyed at the defect
and the energy gain of
the superconducting condensation is lost just locally.
   When the position of the defect coincides with the vortex center,
the loss of the condensation energy is avoided,
because in the vortex core
the superconducting order parameter
is already depleted and therefore the vortex core
could be regarded as a locally realized normal-state region.
   However,
the above argument does not hold
if once we take into account a nonlocal effect
around the defect.
   Thuneberg {\it et al.}~\cite{Thuneberg82,Thuneberg84}
advanced the understanding of the mechanism of
the elementary vortex pinning,
taking into account of such a nonlocal effect
that the defect scatters the quasiparticles
around it as a scattering center.
   They calculated
the vortex pinning energy for
a vortex in isotropic $s$-wave superconductors
(i.e., an isotropic $s$-wave vortex)
and found that the vortex pinning energy is larger than
Eq.\ (\ref{eq:conventional})
by a factor of $\xi/d$, where $\xi$ is the
coherence length
and $d$ the linear dimension of the defect.~\cite{Thuneberg82,Thuneberg84}
   Thuneberg~\cite{Thuneberg84:JLTP} subsequently attacked
the same problem by deriving the impurity (or defect) correction term of
the Ginzburg-Landau (GL) theory for the isotropic $s$-wave superconductors.
   Friesen and Muzikar~\cite{Friesen} extended the GL method of
Ref.~\citen{Thuneberg84:JLTP} to superconductors
with general pairing symmetry
(including the unconventional superconductors).
   Kuli\'c and Dolgov~\cite{Kulic} discussed the vortex pinning potential
due to an anisotropic impurity in the unconventional superconductors.
   The present authors discussed,
in Refs.~\citen{Hayashi02,Hayashi02-2},
a new pinning effect intrinsic to chiral $p$-wave superconductors.

   With these backgrounds, we discuss, in this paper, a relation between
the elementary vortex pinning potential
and the Anderson's theorem for nonmagnetic impurities.
   Here, the Anderson's theorem
means that
nonmagnetic impurities (or nonmagnetic defects)
do not affect the thermodynamic properties
of a superconductor.~\cite{Anderson,Maki,Sigrist91}
   We focus on the following two points.
   (i) The vortex pinning energy in the unconventional superconductors is
enhanced, compared to that in the isotropic $s$-wave superconductors.
   This enhancement originates from the pair-breaking effect
far away from the vortex core in the unconventional superconductors
(i.e., the pair-breaking effect due to
the well-known non-applicability of the Anderson's theorem
in the unconventional superconductors~\cite{Sigrist91}).
   Such a pair-breaking effect
does not occur in the isotropic $s$-wave superconductors
(owing to the applicability of
the Anderson's theorem
in homogeneous isotropic $s$-wave
superconductors~\cite{Anderson,Maki,Sigrist91}).
   The unconventional superconductivity~\cite{Sigrist91} has been proposed
for many superconductors such as high-$T_{\rm c}$ cuprates, organic conductors, and heavy-fermion compounds.
   Thus this issue becomes very important recently.
   (ii) We then consider an effect of the chirality on the vortex pinning
energy in a chiral $p$-wave superconductor
with an unconventional pairing \cite{Sigrist99,Lebed00}
${\bf d}=\bar{\bf z}(\bar{k}_{x} \pm i \bar{k}_{y})$,
which is one of the unconventional superconductors.
   The vortex pinning energy depends on the chirality.
   This is related to local applicability of the Anderson's theorem
(or local recovery of the Anderson's theorem)
inside the vortex core of {\it chiral $p$-wave} superconductors.

   In Sec.\ 2, we summarize the formulation of the quasiclassical theory of
superconductivity used for the study of the vortex pinning.
   In Sec.\ 3, the Anderson's theorem for nonmagnetic impurities
in the isotropic $s$-wave superconductors is
described within the formalism of the quasiclassical theory.
   Even in the vortex states, the Anderson's theorem is applicable
if the nonmagnetic defect is far away from the vortex core.
   By contrast, the nonmagnetic defect inside the vortex core
affects the free energy of superconductors and yields
the vortex pinning energy.
   In Sec.\ 4,
we discuss
the enhancement of the vortex pinning energy
due to the pair-breaking effect of the nonmagnetic defect
far away from the vortex core in the unconventional superconductors.
   In Sec.\ 5,
we discuss the effect of the nonmagnetic defect inside
the vortex core in the chiral $p$-wave superconductor.
   The inside of the chiral $p$-wave vortex core
is similar to the homogeneous state of an isotropic $s$-wave superconductor.
   As an evidence, we show that the nonmagnetic defect inside
the vortex core does not affect
the free energy of the chiral $p$-wave superconductors.
   We regard this result as local applicability of
the Anderson's theorem
{\it inside the vortex core in the chiral $p$-wave superconductors}.
   The summary is given
in Sec.\ 6.

\section{FORMULATION}
   We adopt
the quasiclassical theory of
superconductivity~\cite{Eilenberger,LO,serene}
to investigate the vortex pinning.
   We consider
the quasiclassical Green function
in the absence of the pinning,
%
\begin{equation}
{\hat g}_{\rm imt}(i\omega_n,{\bf r},{\bar{\bf k}})=
-i\pi
\pmatrix{
g_{\rm imt} &
if_{\rm imt} \cr
-if^{\dagger}_{\rm imt} &
-g_{\rm imt} \cr
},
\label{eq:qcg}
\end{equation}
which is the solution of the Eilenberger equation,~\cite{serene}
\begin{equation}
i {\bf v}_{\rm F}({\bar{\bf k}}) \cdot
{\bf \nabla}{\hat g}_{\rm imt}
+ \bigl[ i\omega_n {\hat \tau}_{z}-{\hat \Delta},
{\hat g}_{\rm imt} \bigr]
=0,
\label{eq:eilen}
\end{equation}
where
the superconducting order parameter is
${\hat \Delta}({\bf r},{\bar{\bf k}}) =
\bigl[ ({\hat \tau}_{x} + i {\hat \tau}_{y}) \Delta({\bf r},{\bar{\bf k}})
- ({\hat \tau}_{x} - i {\hat \tau}_{y}) \Delta^*({\bf r},{\bar{\bf k}})
\bigr] /2$
and ${\hat \tau}_{i}$ are the Pauli matrices.
   ${\bf v}_{\rm F}({\bar{\bf k}})$ is the Fermi velocity,
$\omega_n$ is the fermionic Matsubara frequency,
and
the commutator $[{\hat a},{\hat b}]={\hat a}{\hat b}-{\hat b}{\hat a}$.
   The Eilenberger equation
is supplemented by the normalization condition
${\hat g}_{\rm imt}(i\omega_n,{\bf r},{\bar{\bf k}})^2
=-\pi^2{\hat 1}$.~\cite{serene}
   Since we consider, in this paper,
an isolated single vortex in extreme type-II
superconductors where the Ginzburg-Landau parameter $\kappa \gg 1$,
the vector potential
is neglected in Eq. (\ref{eq:eilen}).
   We use units in which $\hbar = k_{\rm B} = 1$.

   In this paper,
the system is assumed to be a two dimensional conduction layer
perpendicular to the magnetic field.
   The vector ${\bf r}=(r\cos\phi,r\sin\phi)$
is the center of mass coordinate.
   The unit vector
${\bar{\bf k}}
=(\cos\theta, \sin\theta)$
represents
the wave number
of relative motion of the Cooper pairs.
   We assume a circular Fermi surface
and ${\bf v}_{\rm F}({\bar{\bf k}})
=v_{\rm F}{\bar{\bf k}}
=(v_{\rm F}\cos\theta,v_{\rm F}\sin\theta)$.

   The effect of the pinning is introduced to
the quasiclassical theory of superconductivity
as follows.~\cite{Thuneberg82,Thuneberg84,Thuneberg81}
   The quasiclassical Green function ${\hat g}$
in the presence of a point-like nonmagnetic defect situated at
${\bf r}={\bf R}$
is obtained from the Eilenberger equation
\begin{eqnarray}
i {\bf v}_{\rm F}({\bar{\bf k}}) \cdot
{\bf \nabla}{\hat g}
+ \bigl[ i\omega_n {\hat \tau}_{z}-{\hat \Delta}, {\hat g} \bigr]
=\bigl[ {\hat t}, {\hat g}_{\rm imt} \bigr] \delta ({\bf r}'),
\label{eq:eilen-pin}
\end{eqnarray}
and the $t$ matrix due to the nonmagnetic defect
(or nonmagnetic impurity)
\begin{eqnarray}
{\hat t}(i\omega_n, {\bf r}') =
\frac{v}{D} \Bigl[
{\hat 1} + N_{0} v
\langle {\hat g}_{\rm imt}(i\omega_n, {\bf r}',{\bar {\bf k}}) \rangle_\theta
\Bigr],
\label{eq:t-matrix}
\end{eqnarray}
where ${\bf r}'={\bf r}-{\bf R}$,
the denominator $D=1+(\pi N_0 v)^2 \bigl[
\langle g_{\rm imt} \rangle_\theta ^2
+ \langle f_{\rm imt} \rangle_\theta
\langle f^{\dagger}_{\rm imt} \rangle_\theta
\bigr]$,
the average over the Fermi surface
$\langle \cdots \rangle_\theta = \int \cdots d \theta/2\pi$,
the normal-state density of states on the Fermi surface $N_{0}$,
and
we assume the $s$-wave scattering $v$
when obtaining Eq.\ (\ref{eq:t-matrix}).
   We define a parameter
$\sin^2 \delta_0 = (\pi N_0 v)^2 /
\bigl[ 1+ (\pi N_0 v)^2 \bigr]$,
which measures how strong
the scattering potential of
the nonmagnetic defect is.

   The vortex pinning potential $\delta \Omega (R)$
($ R \equiv |{\bf R}|$),
i.e.,
the difference
in the free energy
between the states with and without the nonmagnetic defect
is,
at the temperature $T$, given
as~\cite{Thuneberg82,Thuneberg84,Thuneberg81,Viljas}
\begin{eqnarray}
\delta \Omega (R)
&=& \Omega_p (R) - \Omega_0 (R) \nonumber\\
&=& N_0 T \int^{1}_{0} d \lambda
\sum_{\omega_{n}=-\infty}^{\infty}
\int \frac{d \theta}{2\pi}
\int d {\bf r}
{\rm Tr}
\bigl[
\delta {\hat g}_{\lambda} {\hat \Delta}_b
\bigr],
\label{eq:free-ene2}
\end{eqnarray}
where
$\Omega_p (R)$ is the free energy in the {\it presence} of
the nonmagnetic defect,
$\Omega_0 (R)$ the free energy in the {\it absence} of
the nonmagnetic defect,
$\delta {\hat g}_{\lambda} = {\hat g}-{\hat g}_{\rm imt}$
is evaluated at
${\hat \Delta}= \lambda {\hat \Delta}_{b}$,
and
${\hat \Delta}_{b}$
is the order parameter in the absence of the nonmagnetic defect.


\section{ELEMENTARY PINNING POTENTIAL FOR ISOTROPIC $S$-WAVE SUPERCONDUCTORS}
   The elementary pinning potential for
the isotropic $s$-wave superconductors has been microscopically calculated
first by Thuneberg {\it et al.}~\cite{Thuneberg82,Thuneberg84}
on the basis of the quasiclassical theory of superconductivity.
   In this section, we summarize their results.
\subsection{Homogeneous System}
   In the homogeneous system
(e.g., far away from the vortex core),
the matrix elements of the solution of
the Eilenberger equation (\ref{eq:eilen}) are
given as~\cite{Thuneberg84,serene,Klein87}
\begin{equation}
g_{\rm imt} = \frac{\omega_n}{\sqrt{\omega_n^2 +|\Delta|^2}}, \quad
f_{\rm imt} = \frac{\Delta}{\sqrt{\omega_n^2 +|\Delta|^2}}, \quad
f^{\dagger}_{\rm imt} = \frac{\Delta^{*}}{\sqrt{\omega_n^2 +|\Delta|^2}},
\label{eq:bulk-gf}
\end{equation}
where $\Delta$ is the spatially uniform order parameter
in an isotropic $s$-wave superconductor.
   Because these have no $\bar{\bf k}$- (or $\theta$-) dependence,
the Fermi-surface averages of them are
\begin{equation}
\langle g_{\rm imt} \rangle_\theta
   =g_{\rm imt}, \quad
\langle f_{\rm imt} \rangle_\theta
   =f_{\rm imt}, \quad
\langle f^{\dagger}_{\rm imt} \rangle_\theta
   =f^{\dagger}_{\rm imt},
\label{eq:bulk-gf-ave}
\end{equation}
namely,
\begin{equation}
\langle \hat{g}_{\rm imt} \rangle_\theta
   =\hat{g}_{\rm imt}.
\label{eq:bulk-g}
\end{equation}
   Inserting this Eq.\ (\ref{eq:bulk-g})
into the impurity $t$ matrix, Eq.\ (\ref{eq:t-matrix}),
we obtain
\begin{equation}
{\hat t}=
\frac{v}{D} \Bigl[
{\hat 1} + N_{0} v
{\hat g}_{\rm imt}
\Bigr],
\label{eq:t-matrix-bulk}
\end{equation}
and therefore the right hand side of Eq.\ (\ref{eq:eilen-pin}) vanishes,
namely,
\begin{equation}
\bigl[ {\hat t}, {\hat g}_{\rm imt} \bigr] =0.
\label{eq:commu-bulk}
\end{equation}
   When $[{\hat t},{\hat g}_{\rm imt}]=0$,
the Eilenberger equation (\ref{eq:eilen-pin})
in the presence of the nonmagnetic defect
is identical to Eq.\ (\ref{eq:eilen})
(the equation in the absence of the nonmagnetic defect),
namely, the impurity has no influence on the Green function and
the free energy
$\bigl($thus, $\delta \Omega= 0$$\bigr)$.
   This is consistent with the Anderson's theorem,
which states that nonmagnetic impurities do not change the free energy
of homogeneous isotropic $s$-wave
superconductors.~\cite{Anderson,Maki,Sigrist91}
   It is noted that the Anderson's theorem is described as
the commutativity
$\bigl($Eq.\ (\ref{eq:commu-bulk})$\bigr)$
between
the impurity $t$-matrix, $\hat t$, and
the intermediate green function,~\cite{Thuneberg84,Thuneberg81}
${\hat g}_{\rm imt}$, within
the framework of
the quasiclassical theory
of superconductivity.

\subsection{Nonmagnetic Impurity inside Vortex Core}
   In the spatially varying case, the Anderson's theorem does not apply.
   Therefore, the nonmagnetic defect within a vortex core affects
the free energy of the system
and has a contribution to the vortex pinning energy.

   At the vortex center ${\bf r}=0$,
on the basis of an analysis of
the so-called zero-core vortex model,
the matrix elements of the solution of Eq.\ (\ref{eq:eilen}) are
approximately given as~\cite{Thuneberg84}
\begin{equation}
g_{\rm imt} = \frac{\sqrt{\omega_n^2 +|{\tilde \Delta}|^2}}{\omega_n}, \quad
f_{\rm imt} = \frac{-{\tilde \Delta}}{\omega_n}, \quad
f^{\dagger}_{\rm imt} = \frac{{\tilde \Delta}^{*}}{\omega_n},
\label{eq:z-core-gf}
\end{equation}
where
${\tilde \Delta}=
\Delta(r \rightarrow \infty)\exp(i \theta)$.
   Here, the vortex with
the order parameter $\Delta({\bf r})=\Delta(r)\exp(i \phi)$
is considered,
and the amplitude of the order parameter $\Delta(r)$
is set to be constant (i.e., zero core) around the vortex,
which is the approximation
based on the zero-core vortex model.~\cite{Thuneberg84}
    When Eq.\ (\ref{eq:z-core-gf}) is obtained,
the quasiparticles
which go through the origin ${\bf r}=0$ are considered.~\cite{Thuneberg84}
   The position vectors of such quasiparticles
are parallel to the Fermi velocity
$\bigl($i.e., ${\bf r} \parallel
{\bf v}_{\rm F}({\bar {\bf k}})$$\bigr)$,
and therefore
$\phi=\theta$ in ${\tilde \Delta}$ of Eq.\ (\ref{eq:z-core-gf}).

   The Fermi-surface averages of Eq.\ (\ref{eq:z-core-gf}) are
\begin{equation}
\langle g_{\rm imt} \rangle_\theta
   =g_{\rm imt}, \quad
\langle f_{\rm imt} \rangle_\theta
   =0, \quad
\langle f^{\dagger}_{\rm imt} \rangle_\theta
   =0,
\label{eq:v-gf-ave}
\end{equation}
namely,
\begin{equation}
\langle \hat{g}_{\rm imt} \rangle_\theta
\neq
\hat{g}_{\rm imt}.
\label{eq:v-g}
\end{equation}
   Because of the phase factor
$\exp(i \theta)$
of ${\tilde \Delta}$
in Eq.\ (\ref{eq:z-core-gf}),
the Fermi-surface averages of
the anomalous Green functions,
$\langle f_{\rm imt} \rangle_\theta$
and $\langle f^{\dagger}_{\rm imt} \rangle_\theta$,
vanish in Eq.\ (\ref{eq:v-gf-ave}).

\begin{figure}
%
\centerline{\psfig{file=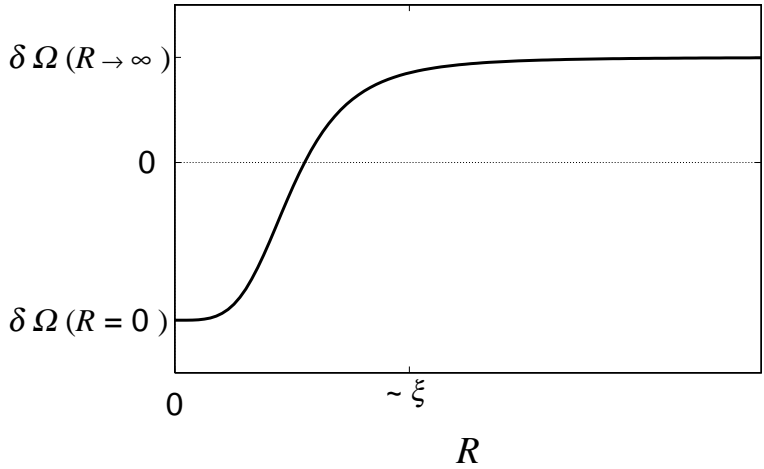,height=8.0cm}}
%
\vspace{-10mm}
\caption{
   Schematic figure of the vortex pinning potential
$\delta \Omega(R)$
as a function of the distance $R$
between the vortex center and the nonmagnetic defect.
   The recovery length of $\delta \Omega(R)$
is of the order of
the coherence length $\xi$.
}
\label{fig:1}
\end{figure}

   From Eqs.\ (\ref{eq:t-matrix}) and (\ref{eq:v-g}),
it follows that
the $t$ matrix due to the nonmagnetic defect situated at the vortex center
does not commute with the Green function ${\hat g}_{\rm imt}$,
namely $[{\hat t},{\hat g}_{\rm imt}] \neq 0$ in general.
   Therefore, a nonzero effect of the nonmagnetic defect
appears in the right hand side of
the Eilenberger equation (\ref{eq:eilen-pin})
and the nonmagnetic defect situated at the vortex center
affects the Green function and the free energy,
$\delta \Omega (R=0)
= \Omega_p (R=0) - \Omega_0 (R=0) <0$.
($R$ is the distance between the vortex center and the nonmagnetic defect.)
   The nonzero $\delta \Omega (R=0)$ is an origin of the vortex pinning
(see Fig.\ \ref{fig:1}).
   In the isotropic $s$-wave
superconductors $\bigl($in which
$\delta \Omega(R \rightarrow \infty) =0$
as discussed in Sec.\ 3.1$\bigr)$,
it is the only origin of the vortex pinning.

   According to Thuneberg {\it et al.},~\cite{Thuneberg82,Thuneberg84}
an approximated
analytical expression for the vortex pinning potential
in the isotropic $s$-wave superconductors
is given as
\begin{equation}
\delta \Omega (R=0)
= - 2 T \ln \cosh
\Bigl[
\frac{|{\tilde \Delta}(T)| \sin \delta_0}{2 T}
\Bigr],
\label{eq:v-pin-potential}
\end{equation}
on the basis of the zero-core vortex model.~\cite{Thuneberg84}
Here, $|{\tilde \Delta}(T)|$ has the temperature dependence of
the BCS gap, $\Delta_{\rm BCS}(T)$.
   From the quantitative viewpoint, however,
it should be noted that
at high temperatures
Eq.\ (\ref{eq:v-pin-potential})
overestimates the magnitude $|\delta \Omega|$ at one order larger value
as compared to
a precise numerical result.~\cite{Thuneberg82,Thuneberg84}

\section{ENHANCEMENT OF PINNING ENERGY IN UNCONVENTIONAL SUPERCONDUCTORS}
   As seen in the preceding section,
in the case of the isotropic $s$-wave superconductors,
the nonmagnetic defect does not affect the vortex pinning potential
$\delta \Omega$ far away from the vortex core
$\bigl($i.e., $\delta \Omega(R \rightarrow \infty) = 0$$\bigr)$,
because
the Anderson's theorem is applicable to
the homogeneous system in the isotropic $s$-wave superconductors.
%
   In the unconventional superconductors, on the other hand,
the Anderson's theorem does not apply
even in the homogeneous system.~\cite{Sigrist91}
This is because the order parameter $\Delta$ in Eq.\ (\ref{eq:bulk-gf})
has a $\bar{\bf k}$- (or $\theta$-) dependence
in such superconductors
and hence
$\langle \hat{g}_{\rm imt} \rangle_\theta
   \neq \hat{g}_{\rm imt}$
instead of Eq.\ (\ref{eq:bulk-g}).
   Therefore, the nonmagnetic defect affects the free energy
and gives rise to
the condensation energy loss,
$\delta \Omega
= \Omega_p - \Omega_0 >0$,
in the unconventional superconductors.
   Such a condensation energy loss
far away from the vortex core, $\delta\Omega (R \rightarrow \infty)$,
contributes to
the depth of the vortex pinning potential (see Fig.\ \ref{fig:1}),
namely, to the vortex pinning
energy,~\cite{Thuneberg81,Friesen,Kulic,Hayashi02}
\begin{equation}
E_{\rm pin} =
\delta\Omega(R \rightarrow \infty) -
\delta\Omega(R=0).
\label{eq:pin-ene}
\end{equation}
   While
the vortex pinning energy is just $E_{\rm pin}=-\delta\Omega(R=0)$
in the isotropic $s$-wave superconductors,
in the case of the unconventional superconductors
the additional contribution from $\delta\Omega(R \rightarrow \infty)$
may enhance the vortex pinning energy $E_{\rm pin}$.
   In what follows,
from Eq.\ (\ref{eq:free-ene2})
we calculate
$\delta\Omega (R \rightarrow \infty)$
for the chiral $p$-wave superconductor
as an example of the unconventional superconductors.

   For the chiral $p$-wave pairing state~\cite{Sigrist99,Lebed00}
${\bf d}=\bar{\bf z}({\bar k}_x \pm i{\bar k}_y)
= \bar{\bf z} \exp(\pm i \theta)$,
we consider the condensation energy loss
when putting a nonmagnetic defect on
the homogeneous system.
   The spatially uniform order parameter in the homogeneous system
is expressed as
$\Delta_b({\bar {\bf k}})=\Delta_b(\theta)=\Delta_b^0\exp(\pm i \theta)$.
   The matrix elements of the solution of Eq.\ (\ref{eq:eilen}) are
given by
\begin{equation}
g_{\rm imt}
 = \frac{\omega_n}{\sqrt{\omega_n^2 +|\Delta_b^0|^2}}, \quad
f_{\rm imt}
 = \frac{\Delta_b(\theta)}{\sqrt{\omega_n^2 +|\Delta_b^0|^2}}, \quad
f^{\dagger}_{\rm imt}
 = \frac{\Delta^{*}_b(\theta)}{\sqrt{\omega_n^2 +|\Delta_b^0|^2}},
\label{eq:bulk-gf-p}
\end{equation}
as in Eq.\ (\ref{eq:bulk-gf}).
   The Fermi-surface averages of Eq.\ (\ref{eq:bulk-gf-p}) are
\begin{equation}
\langle g_{\rm imt} \rangle_\theta
   =g_{\rm imt}, \quad
\langle f_{\rm imt} \rangle_\theta
   =0, \quad
\langle f^{\dagger}_{\rm imt} \rangle_\theta
   =0,
\label{eq:bulk-gf-p-ave}
\end{equation}
because $\langle \Delta_b(\theta) \rangle_\theta = 0$.
   Inserting these into Eq.\ (\ref{eq:t-matrix}),
we obtain the commutator $[\hat{t},\hat{g}_{\rm imt}]$
as
\begin{equation}
[\hat{t},\hat{g}_{\rm imt}]=
\frac{A}{2}
\Bigl[
(\hat{\tau}_x + i \hat{\tau}_y) \Delta
+ (\hat{\tau}_x - i \hat{\tau}_y) \Delta^{*}
\Bigr],
\label{eq:commutator}
\end{equation}
   where
\begin{equation}
A =
\frac{-2i\omega_n \sin^2 \delta_0}
{N_0 (\omega_n^2 + |\Delta|^2 \cos^2 \delta_0)},
\label{eq:A}
\end{equation}
and $\Delta \equiv \lambda \Delta_b(\theta)$.

   We take a coordinate system with the origin at the nonmagnetic defect:
${\bf r}=s {\bf v}_{\rm F}(\bar{\bf k})/v_{\rm F}
+ b
\bigl[\bar{\bf z} \times {\bf v}_{\rm F}(\bar{\bf k})/v_{\rm F} \bigr]
=s \bar{\bf k} + b (\bar{\bf z} \times \bar{\bf k})$.
   In this coordinate system,
$i {\bf v}_{\rm F}({\bar{\bf k}}) \cdot {\bf \nabla}
= i v_{\rm F} d/d s$
and the Eilenberger equation (\ref{eq:eilen-pin})
is written as
\begin{eqnarray}
i v_{\rm F} \frac{d}{d s} {\hat g}
+ \bigl[ i\omega_n {\hat \tau}_{z}-{\hat \Delta}, {\hat g} \bigr]
=\bigl[ {\hat t}, {\hat g}_{\rm imt} \bigr] \delta (s) \delta(b),
\label{eq:eilen-pin-bulk}
\end{eqnarray}
and $\delta \hat{g}_\lambda$ ($= \hat{g} - \hat{g}_{\rm imt}$)
in Eq.\ (\ref{eq:free-ene2})
follows the equation,
\begin{eqnarray}
i v_{\rm F} \frac{d}{d s} \delta{\hat g}_\lambda
+ \bigl[ i\omega_n {\hat \tau}_{z}-{\hat \Delta},
\delta{\hat g}_\lambda \bigr]
=\bigl[ {\hat t}, {\hat g}_{\rm imt} \bigr] \delta (s),
\label{eq:eilen-pin-bulk-gl}
\end{eqnarray}
where $\delta \hat{g}_\lambda = 0$ for $b \neq 0$.

   Using the Fourier transformations,
$\delta \hat{g}_\lambda (s) =
\int dq \exp(iqs) \delta \hat{g}_\lambda (q) /2\pi$,
$\delta (s) = \int dq \exp(iqs) /2\pi$,
and a decomposition
$\delta \hat{g}_\lambda (s) =
g_1 \hat{\tau}_x +g_2 \hat{\tau}_y +g_3 \hat{\tau}_z$,
we can solve Eq.\ (\ref{eq:eilen-pin-bulk-gl})
into which Eq.\ (\ref{eq:commutator}) is inserted.
   From its solution $\delta \hat{g}_\lambda (q)$, we obtain
\begin{eqnarray}
\int d{\bf r} \delta \hat{g}_\lambda
   &=& \int ds \delta \hat{g}_\lambda (s) \\ \nonumber
   &=& \delta \hat{g}_\lambda (q=0) \\ \nonumber
   &=& \frac{A \omega_n
\bigl[ -i(\Delta-\Delta^*)\hat{\tau}_x
        +(\Delta+\Delta^*)\hat{\tau}_y
        +(4\omega_n^2 + 2|\Delta|^2)\hat{\tau}_z \bigr] }
            {4(\omega_n^2 + |\Delta|^2)},
\label{eq:int-g}
\end{eqnarray}
and therefore, from Eq.\ (\ref{eq:free-ene2}),
\begin{eqnarray}
\delta \Omega
&=& N_0 T \int^{1}_{0} d \lambda       \nonumber
\sum_{\omega_{n}}
\int \frac{d \theta}{2\pi}
\frac{2\omega_n^2 \lambda |\Delta_b|^2 \sin^2 \delta_0}
 {N_0 (\omega_n^2 + \lambda^2 |\Delta_b|^2 \cos^2 \delta_0)
  (\omega_n^2 + \lambda^2 |\Delta_b|^2)} \\
&=& 2T \ln
\biggl[
\frac{\cosh \bigl(|\Delta_b^0(T)| / 2T \bigr)}
{\cosh \bigl(|\Delta_b^0(T)|\cos \delta_0 /2T \bigr)}
\biggr].
\label{eq:ene-loss}
\end{eqnarray}
   This is the condensation energy loss due to
the pair-breaking effect,
$\delta\Omega (R \rightarrow \infty)$,
for the chiral $p$-wave superconductor.
Here, $|\Delta_b^0(T)|$ has the temperature dependence of
the BCS gap, $\Delta_{\rm BCS}(T)$.


   At high temperatures,
the value $\delta\Omega (R \rightarrow \infty)$ of
Eq.\ (\ref{eq:ene-loss})
becomes the same order as the magnitude
$|\delta\Omega (R=0)|$
of Eq.\ (\ref{eq:v-pin-potential}).
   Therefore,
it appears that
at high temperatures
the vortex pinning energy
in the unconventional superconductors,
$E_{\rm pin}$
$\bigl(=\delta\Omega (R \rightarrow \infty)
+|\delta \Omega (R=0)| \bigr)$,
is enhanced about twice
compared to the isotropic $s$-wave superconductor
in which $E_{\rm pin}=|\delta \Omega (R=0)|$,
owing to the additional contribution
from $\delta\Omega (R \rightarrow \infty)$.
   However, since
the approximated expression, Eq.\ (\ref{eq:v-pin-potential}),
overestimates the magnitude $|\delta \Omega (R=0)|$
at one order larger value
at high temperatures,~\cite{Thuneberg82,Thuneberg84}
we may expect that
the value of $\delta\Omega (R \rightarrow \infty)$ is two or more times
larger than an actual value of $|\delta \Omega (R=0)|$.
   Indeed we have found, using a numerical calculation,
that at high temperatures
the vortex pinning energy $E_{\rm pin}$
in the case of
the chiral $p$-wave superconductor~\cite{Hayashi02-2}
(and a $d$-wave superconductor~\cite{Hayashi02})
is
about 10 times larger than that of the isotropic $s$-wave
pairing case,
owing to $\delta\Omega (R \rightarrow \infty)
\sim 10 \times |\delta \Omega (R=0)|$.~\cite{Hayashi02,Hayashi02-2}
   This enhancement of the vortex pinning energy $E_{\rm pin}$
is due to the pair-breaking effect of
the nonmagnetic defect far away from the vortex center
 and therefore such an enhancement is a common feature of
the unconventional superconductors.

\section{ANDERSON'S THEOREM INSIDE VORTEX CORE IN CHIRAL SUPERCONDUCTOR}
   As described in Sec.\ 3,
the Anderson's theorem does not apply to spatially varying system
(e.g., the vortex core)
and therefore the nonmagnetic defect
inside the vortex core affects the free energy
$\delta \Omega(R=0)$
and acts as a pinning center.
   In general, it holds also for the unconventional superconductors.
   However, an exception exists.
   In this section, we discuss novel applicability of
the Anderson's theorem inside the vortex core
in the chiral $p$-wave superconductors;
for those superconductors,
the nonmagnetic defect
inside a vortex core does not contribute to the free energy
and, as a result, the vortex pinning potential at the vortex center turns
out to be $\delta \Omega(R=0) =0$.
   This phenomenon originates from a quantum effect,
i.e., a cancellation of the phase factors of
the superconducting order parameter.

   For the chiral $p$-wave pairing state~\cite{Sigrist99,Lebed00}
${\bf d}=\bar{\bf z}({\bar k}_x \pm i{\bar k}_y)
= \bar{\bf z} \exp(\pm i \theta)$,
it is known that
the order parameter around a single vortex
$\Delta({\bf r},{\bar {\bf k}})$
$\bigl[\equiv \Delta(r,\phi;\theta) \bigr]$
has two possible forms
depending on whether
the chirality and vorticity are parallel or
antiparallel each other.~\cite{Heeb99,Matsumoto01,Kato01}
%
   One form is
\begin{eqnarray}
\Delta^{+-}(r,\phi;\theta)=
\Delta_{+}(r) e^{i(\theta-\phi)}
+ \Delta_{-}(r) e^{i(-\theta+\phi)},
\label{eq:op-pm}
\end{eqnarray}
where the chirality and vorticity are antiparallel
$\bigl[$$p(+-)$-case$\bigr]$.
   The other is
\begin{eqnarray}
\Delta^{++}(r,\phi;\theta)=
\Delta_{+}(r) e^{i(\theta+\phi)}
+ \Delta_{-}(r) e^{i(-\theta+3\phi)},
\label{eq:op-pp}
\end{eqnarray}
where the chirality and vorticity are parallel
$\bigl[$$p(++)$-case$\bigr]$.
   Here, the vortex center is situated at ${\bf r}=0$,
the dominant component
$\Delta_{+}(r \rightarrow \infty)=\Delta_{\rm BCS}(T)$,
and
the induced one
$\Delta_{-}(r \rightarrow \infty)=0$.
   Because of axisymmetry of the system,
we can take
$\Delta_{\pm}(r)$
to be real.

   From the quasiclassical viewpoint,
the quasiparticles inside a vortex core
run along straight lines called as
quasiparticle paths.\cite{Klein89,Rainer,Hayashi97}
   We consider the quasiparticle paths
which go through the origin ${\bf r}=0$.
   On those paths,
the position vector is parallel to the direction of the quasiparticle path
(i.e., ${\bf r} \parallel {\bar {\bf k}}$), and therefore
$\phi=\theta,\ \theta+\pi$.
   In this situation ($\phi=\theta$),
from Eqs.\ (\ref{eq:op-pm}) and (\ref{eq:op-pp}),
the order parameter on the path is
\begin{eqnarray}
\Delta^{+-}(r, \phi=\theta; \theta) =
\Delta_{+}(r) + \Delta_{-}(r)
\label{eq:op-pm-c}
\end{eqnarray}
in $p(+-)$-case,
and
\begin{eqnarray}
\Delta^{++}(r, \phi=\theta; \theta) =
\bigl[ \Delta_{+}(r) + \Delta_{-}(r) \bigr] e^{2i\theta}
\label{eq:op-pp-c}
\end{eqnarray}
in $p(++)$-case.
   The cancellation between
the phase factors,
$\exp(i\theta)$  due to the chirality of the Cooper pair
and $\exp(i\phi)$ due to the vorticity of the vortex,
occurs in Eq.\ (\ref{eq:op-pm-c})
and not in Eq.\ (\ref{eq:op-pp-c}).

   On the basis of the zero-core vortex model,~\cite{Thuneberg84}
as in Eq.\ (\ref{eq:z-core-gf})
the matrix elements of
${\hat g}_{\rm imt}$ at the vortex center are
approximately obtained as
\begin{equation}
g_{\rm imt} = \frac{\sqrt{\omega_n^2 +|{\tilde \Delta}|^2}}{\omega_n}, \quad
f_{\rm imt} = \frac{-{\tilde \Delta}}{\omega_n}, \quad
f^{\dagger}_{\rm imt} = \frac{{\tilde \Delta}^{*}}{\omega_n},
\label{eq:z-core-gf-p}
\end{equation}
where
${\tilde \Delta}=
\Delta^{+\pm}(r \rightarrow \infty, \phi=\theta; \theta)$.
   Inserting the order parameter of Eq.\ (\ref{eq:op-pm-c})
into Eq.\ (\ref{eq:z-core-gf-p}),
we obtain
the Green function
integrated over the Fermi surface as,
in $p(+-)$-case,
\begin{equation}
\langle {\hat g}_{\rm imt} \rangle_\theta = {\hat g}_{\rm imt}, \quad
\langle f_{\rm imt} \rangle_\theta = f_{\rm imt}, \quad
\langle f_{\rm imt}^{\dagger} \rangle_\theta = f_{\rm imt}^{\dagger},
\label{eq:f-caseI}
\end{equation}
namely,
\begin{equation}
\langle \hat{g}_{\rm imt} \rangle_\theta
=
\hat{g}_{\rm imt},
\label{eq:v-g-p-I}
\end{equation}
because of the absence of any phase factors in Eq.\ (\ref{eq:op-pm-c}),
i.e., because the cancellation between the chirality factor
$\exp(i \theta)$
and the vorticity factor
$\exp(-i \phi)$ occurs
and then $\langle \Delta^{+-} \rangle_\theta = \Delta^{+-}$.
   We obtain
$[{\hat t},{\hat g}_{\rm imt}]=0$
from Eqs.\ (\ref{eq:t-matrix}) and (\ref{eq:v-g-p-I}).
   Therefore, following the same discussion as in Sec.\ 3.1,
we conclude that
in this $p(+-)$-case
the nonmagnetic defect at the vortex center
does not affect the free energy and $\delta \Omega(R=0) = 0$.
   This situation is the same as
that of the homogeneous system in the isotropic $s$-wave superconductors
(see Sec.\ 3.1).
   That is, the Anderson's theorem is applicable
(or is recovered)
locally at the vortex center
in $p(+-)$-case when the chirality is antiparallel to the vorticity.

   On the other hand,
in $p(++)$-case,
\begin{equation}
\langle {\hat g}_{\rm imt} \rangle_\theta = {\hat g}_{\rm imt}, \quad
\langle f_{\rm imt} \rangle_\theta = 0, \quad
\langle f_{\rm imt}^{\dagger} \rangle_\theta = 0,
\label{eq:f-caseII}
\end{equation}
namely,
\begin{equation}
\langle \hat{g}_{\rm imt} \rangle_\theta
\neq
\hat{g}_{\rm imt},
\label{eq:v-g-p-II}
\end{equation}
because the phase factor $\exp(2i\theta)$ is contained in
Eq.\ (\ref{eq:op-pp-c}) and then
$\langle \Delta^{++} \rangle_\theta$ =0.
   Therefore, $[{\hat t},{\hat g}_{\rm imt}] \neq 0$
from Eqs.\ (\ref{eq:t-matrix}) and (\ref{eq:v-g-p-II}),
and $\delta \Omega(0) \neq 0$
in this $p(++)$-case
when the sense of the chirality is the same as that of the vorticity.
   This is the same as the impurity effect inside the vortex core discussed
in Sec.\ 3.2.

   The above analysis in this section
is based on the zero-core vortex model,
i.e., on the non-self-consistent (constant) amplitude of
the order parameter,
but the phase of the order parameter,
which is important for the above results,
is correctly taken into account by that zero-core vortex model.
   Therefore, the essential physics is captured by the above analysis.
   We have certainly confirmed it
by a numerical calculation of $\delta \Omega(R)$
based on self-consistently obtained
order parameter.~\cite{Hayashi02-2} Related phenomena have been discussed
in Refs.~\citen{Volovik,MS,Kato00,Kato02} in different contexts.

\begin{table}
\caption{
   The Anderson's theorem is applicable
($\bigcirc$)
or {\it not} applicable ($\times$)
at the vortex center ($R=0$)
and far away from the vortex core ($R \rightarrow \infty$)
in certain superconductors.
If applicable ($\bigcirc$),
the vortex pinning potential $\delta \Omega$ becomes zero there.
Refer also to Fig.\ 1.
}
\begin{center}
\begin{tabular}{lcc}
Superconductors &  $R=0$    &  $R \rightarrow \infty$  \\
\hline
 $s$-wave                   & $\times$   & $\bigcirc$   \\
unconventional [general]    & $\times$   & $\times$   \\
chiral $p$-wave $\bigl[$$p(+-)$-case$\bigr]$
                            & $\bigcirc$ & $\times$
\end{tabular}
\end{center}
\label{table:1}
\end{table}

\section{SUMMARY}
   We have discussed the relation between
the vortex pinning energy and the Anderson's theorem,
which is summarized in Table 1.
   The vortex pinning energy,
$E_{\rm pin}$ $\bigl($Eq.\ (\ref{eq:pin-ene})$\bigr)$,
is given by the difference
in the free energy
between the case when vortex is
far away from the nonmagnetic defect
$\delta \Omega(R\rightarrow \infty)$
and
the case when the vortex is located at the nonmagnetic defect
$\delta \Omega(R=0)$.
   The nonmagnetic defect far away from the vortex core can reduce locally
the condensation energy of superconductors.
   On the other hand, the nonmagnetic defect within the vortex core can yield
the energy gain
to superconductors (through the scattering of quasiparticles
in zero energy bound state).
   These two factors,
$\delta \Omega(R\rightarrow \infty)$ and
$\delta \Omega(R=0)$,
determine the magnitude of the vortex pinning energy $E_{\rm pin}$
for superconductors (Fig.\ 1).
   However, one of these factors happens to vanish
in the following two cases.
   (1) In the isotropic $s$-wave superconductors,
the nonmagnetic defect far away from the vortex core does not change
the free energy, i.e.,
$\delta \Omega(R\rightarrow \infty)=0$;
this is well known as
the Anderson's theorem.~\cite{Anderson,Maki,Sigrist91}
   (2) The nonmagnetic defect
inside the chiral $p$-wave vortex core,
does not change
the free energy and hence does not yield the energy gain to
superconductors, i.e.,
$\delta \Omega(R=0)=0$,
if the total angular momentum or (equivalently)
the sum of the vorticity and the chirality is zero
$\bigl($$p(+-)$-case in Sec.\ 5$\bigr)$.
   The vanishing angular momentum
$\bigl($or the resulting absence of the phase factors in
the order parameter of Eq.\ (\ref{eq:op-pm-c})$\bigr)$
makes
the chiral $p$-wave vortex core similar
to bulk $s$-wave superconductors.
   Therefore, the absence of the impurity effect in
the chiral $p$-wave vortex core
can be regarded as a consequence of
the Anderson's theorem.
   This is the local applicability of the Anderson's theorem
(or the local recovery of the Anderson's theorem)
inside the chiral $p$-wave vortex core.

   If we compare the isotropic $s$-wave superconductor
with the (generic) unconventional
superconductors in Table 1,
the vortex pinning energy of the latter is larger than
that of the former by the loss of
the condensation energy at $R=\infty$.
   In high-$T_{\rm c}$ cuprates,
this enhancement might be one of the reasons why small
defects such as oxygen vacancies~\cite{Kes,Kes2} and Zn atoms~\cite{Pan}
are efficient pinning centers,
because the high-$T_{\rm c}$ cuprates are believed
to be $d$-wave superconductors
(i.e., the unconventional superconductors).
 On the other hand, the chirality dependence of the vortex pinning
is expected
to be experimentally observed in future, for example,
in
a superconducting material Sr$_2$RuO$_4$.~\cite{Sigrist99,Lebed00,Maeno}





\begin{thebibliography}{9}

\bibitem{Thuneberg82}
E. V. Thuneberg, J. Kurkij\"arvi, and D. Rainer,
{\it Phys. Rev. Lett.} {\bf 48}, 1853 (1982).

\bibitem{Thuneberg84}
E. V. Thuneberg, J. Kurkij\"arvi, and D. Rainer,
{\it Phys. Rev. B} {\bf 29}, 3913 (1984).

\bibitem{Thuneberg84:JLTP}
E. V. Thuneberg,
{\it J. Low Temp. Phys.} {\bf 57}, 415 (1984).

\bibitem{Kerchner83}
H. R. Kerchner, D. K. Christen, C. E. Klabunde,
S. T. Sekula, and R. R. Coltman, Jr.,
{\it Phys. Rev. B} {\bf 27}, 5467 (1983).

\bibitem{Kes}
T. W. Li, A. A. Menovsky, J. J. M. Franse, and P. H. Kes,
{\it Physica C} {\bf 257}, 179 (1996).

\bibitem{Friesen}
M. Friesen and P. Muzikar,
{\it Phys. Rev. B} {\bf 53}, R11953 (1996);
{\it Phys. Rev. B} {\bf 55}, 509 (1997);
{\it Phys. Rev. B} {\bf 57}, 2709 (1998).

\bibitem{Kulic}
M. L. Kuli\'c and O. V. Dolgov,
{\it Phys. Rev. B} {\bf 60}, 13062 (1999).

\bibitem{Hayashi02}
N. Hayashi and Y. Kato,
{\it Physica C} {\bf 367}, 41 (2002).

\bibitem{Hayashi02-2}
N. Hayashi and Y. Kato,
{\it Phys. Rev. B} {\bf 66}, 132511 (2002).

\bibitem{Anderson}
P. W. Anderson,
{\it J. Phys. Chem. Solids} {\bf 11}, 26 (1959).

\bibitem{Maki}
K. Maki,
in {\it Superconductivity} Vol. 2,
edited by R. D. Parks
(Marcel Dekker, N.Y., 1969),
Chap. 18, p. 1035 and p. 1041.

\bibitem{Sigrist91}
M. Sigrist and K. Ueda,
{\it Rev. Mod. Phys.} {\bf 63}, 239 (1991).

\bibitem{Sigrist99}
M. Sigrist, D. Agterberg, A. Furusaki, C. Honerkamp,
K. K. Ng, T. M. Rice, and M. E. Zhitomirsky,
{\it Physica C} {\bf 317-318}, 134 (1999).

\bibitem{Lebed00}
A. G. Lebed and N. Hayashi,
{\it Physica C} {\bf 341-348}, 1677 (2000).

\bibitem{Eilenberger}
G. Eilenberger,
{\it Z. Phys.} {\bf 214}, 195 (1968).

\bibitem{LO}
A. I. Larkin and Yu. N. Ovchinnikov,
{\it Zh. \'Eksp. Teor. Fiz.} {\bf 55}, 2262 (1968)
[{\it Sov. Phys. JETP} {\bf 28}, 1200 (1969)].

\bibitem{serene}
J. W. Serene and D. Rainer,
{\it Phys. Rep.} {\bf 101}, 221 (1983).

\bibitem{Thuneberg81}
E. V. Thuneberg, J. Kurkij\"arvi, and D. Rainer,
{\it J. Phys. C: Solid State Phys.} {\bf 14}, 5615 (1981).

\bibitem{Viljas}
J. K. Viljas and E. V. Thuneberg,
{\it Phys. Rev. B} {\bf 65}, 064530 (2002).
In this reference, the expression conrresponding to
Eq.\ (\ref{eq:free-ene2})
but free from the ${\bf r}$ integration has been derived.

\bibitem{Klein87}
U. Klein,
{\it J. Low Temp. Phys.} {\bf 69}, 1 (1987).

\bibitem{Heeb99}
R. Heeb and D. F. Agterberg,
{\it Phys. Rev. B} {\bf 59}, 7076 (1999).

\bibitem{Matsumoto01}
M. Matsumoto and R. Heeb,
{\it Phys. Rev. B} {\bf 65}, 014504 (2002).

\bibitem{Kato01}
Y. Kato and N. Hayashi,
{\it J. Phys. Soc. Jpn.} {\bf 70}, 3368 (2001).

\bibitem{Klein89}
U. Klein,
{\it Phys. Rev. B} {\bf 40}, 6601 (1989).

\bibitem{Rainer}
D. Rainer, J. A. Sauls, and D. Waxman,
{\it Phys. Rev. B} {\bf 54}, 10094 (1996).

\bibitem{Hayashi97}
N. Hayashi, M. Ichioka, and K. Machida,
{\it Phys. Rev. B} {\bf 56}, 9052 (1997).

\bibitem{Volovik}
G. E. Volovik,
{\it Pis'ma Zh. \'Eksp. Teor. Fiz.} {\bf 70}, 601 (1999)
[{\it JETP Lett.} {\bf 70}, 609 (1999)].

\bibitem{MS}
M. Matsumoto and M. Sigrist,
{\it Physica B} {\bf 281-282}, 973 (2000).

\bibitem{Kato00}
Y. Kato,
{\it J. Phys. Soc. Jpn.} {\bf 69}, 3378 (2000).

\bibitem{Kato02}
Y. Kato and N. Hayashi,
{\it J. Phys. Soc. Jpn.} {\bf 71}, 1721 (2002).

\bibitem{Kes2}
C. J. van der Beek and P. H. Kes,
{\it Phys. Rev. B} {\bf 43}, 13032 (1991).

\bibitem{Pan}
S. H. Pan, E. W. Hudson, A. K. Gupta, K.-W. Ng, H. Eisaki, S. Uchida,
and J. C. Davis,
{\it Phys. Rev. Lett.} {\bf 85}, 1536 (2000).

\bibitem{Maeno}
Y. Maeno, H. Hashimoto, K. Yoshida, S. Nishizaki, T. Fujita,
J. G. Bednorz, and F. Lichtenberg,
{\it Nature} {\bf 372}, 532 (1994).


\end{thebibliography}
\end{document}